\documentclass[11pt, draftclsnofoot, onecolumn,romanappendices]{IEEEtran}
\usepackage[numbers, square, comma, sort&compress]{natbib}
\usepackage{hyperref}\bibfont\small

\usepackage{graphicx, amsfonts,amsmath, amssymb, array,url}
\usepackage{subfigure,setspace,multirow}
\usepackage{epstopdf}
\usepackage{epsfig}

\usepackage[noadjust]{cite}
\usepackage{atbegshi}
\AtBeginDocument{\AtBeginShipoutNext{\AtBeginShipoutDiscard}}

\begin{document}

\title{Performance Analysis of Multi-Antenna Relay Networks over Nakagami-$m$ Fading Channel}
\author{Ehsan Soleimani-Nasab, Ashkan Kalantari and Mehrdad Ardebilipour
\thanks{Authors are with the Faculty of Electrical and Computer Engineering, K.N. Toosi University of Technology, Tehran 1431714191, Iran (e-mail: \{ehsan.soleimani,a.kalantari\}@ee.kntu.ac.ir,  mehrdad@eetd.kntu.ac.ir).}}.

\maketitle
\makeatletter
\long\def\@makecaption#1#2{\ifx\@captype\@IEEEtablestring%
\footnotesize\begin{center}{\normalfont\footnotesize #1}\\
{\normalfont\footnotesize\scshape #2}\end{center}%
\@IEEEtablecaptionsepspace
\else
\@IEEEfigurecaptionsepspace
\setbox\@tempboxa\hbox{\normalfont\footnotesize {#1.}~~ #2}%
\ifdim \wd\@tempboxa >\hsize%
\setbox\@tempboxa\hbox{\normalfont\footnotesize {#1.}~~ }%
\parbox[t]{\hsize}{\normalfont\footnotesize \noindent\unhbox\@tempboxa#2}%
\else
\hbox to\hsize{\normalfont\footnotesize\hfil\box\@tempboxa\hfil}\fi\fi}
\makeatother

\begin{abstract}
In this chapter, the authors present the performance of multi-antenna selective combining decode-and-forward
(SC-DF) relay networks over independent and identically distributed (i.i.d) Nakagami-$m$ fading
channels. The outage probability, moment generation function, symbol error probability and average
channel capacity are derived in closed-form using the SNR statistical characteristics. After that, the
authors formulate the outage probability problem and optimize it with an approximated problem and then
solve it analytically. Finally, for comparison with analytical formulas, the authors perform some Monte-
Carlo simulations.
\end{abstract}
\thispagestyle{empty}


\newpage
\section{Introduction}
Cooperative communication has been an interesting topic for researchers in recent years. Cooperative
communications refer to systems or techniques that allow users to help transmit each other's messages to
the destination. Most cooperative transmission schemes involve two phases of transmission: a
coordination phase, where users exchange their own source data and control messages with each other
and/or the destination, and a cooperation phase, where the users cooperatively retransmit their messages
to the destination.
To enable cooperation among users, different relay technology can be employed depending on the relative
user location, channel condition and transceiver complexity. There are some of the basic cooperative
relaying techniques such as decode-and-forward (DF) in which the relay decodes the received signals and
forwards it either as is or re-encoded to the destination regardless whether the relay can decode correctly
or not.
The relay could also only forward the correctly decoded messages, which is referred to as the selective
DF (S-DF) protocol, amplify-and forward (AF) in which the signal received by the relay is amplified,
frequency translated and retransmitted, coded cooperation (CC) that can be viewed as a generalization of
DF relaying schemes where more powerful channel codes (other than simple repetition codes used in the
DF schemes) are utilized in both phases of the cooperative transmission. When using repetition codes, the
same codeword is transmitted twice (either by the source or the relay) and, thus, bandwidth efficiency is
decreased by one half and compress and forward (CF) schemes refer which to cases where the relay
forwards quantized, estimated, or compressed versions of its observation to the destination. In contrast to
DF or CC schemes, the relay in CF schemes need not decode perfectly the source message but need only
to extract, from its observation, the information that is most relevant to the decoding at the destination.
The amount of information extracted and forwarded to the destination depends on the capacity of the relay destination
link \citep{Dohler}.
\section{Literature Review}
Because the relay selection schemes make an efficient use of time and frequency resources, several selective combining schemes have been introduced in recent years. In \citep{Bletsas}, the authors introduced an opportunistic relaying method which a single relay based on the best end-to-end instantaneous SNR criterion is selected and then forwards the message to the destination. They derived analytical results at high SNRs and the outage probability wasn’t derived in closed-form. The authors in \citep{Norman} and \citep{Hu} analyzed an adaptive DF relay method which only a number of relays were selected to send the messages to the destination. They proved that increasing the number of relays could not always decrease the outage probability.
A selection combiner at the destination with AF relays have been studied in \citep{Sagias:Sigletter} on Nakagami-$m$ fading channels where, a closed-form formula for the outage probability was derived. In \citep{Duong:2009:Comletter}, the authors presented closed-form formulas for the performance of selective DF relaying in Nakagami-$m$ fading channels without considering the direct link between source and destination. The authors in \citep{Ikki} introduced a closed-form expression for the outage probability and average channel capacity using the best relay selection scheme over independent and non-identical Rayleigh fading channels.
However, to the best of our knowledge, no one derived exact closed-form expressions for the outage probability, symbol error probability and average channel capacity for multi-antenna SC-DF relay networks over Nakagami-$m$ fading channels.

 In this chapter, we derived closed-form formulas for outage probability, moment generation function, symbol error probability and average channel capacity. In addition, we minimized the outage probability with optimal and adaptive power allocation. The reminder of this paper is organized as follows. Section II introduces the system model under consideration. Section III gives an analytical approach to evaluate the outage probability, MGF, symbol error probability and average channel capacity of the system. In Section IV, we formulate outage probability problem to optimize the approximate version of our problem which leads to a closed from analytical solution. Finally, Section V presents Monte-Carlo simulation to verify the analytical results.

\section{System Model}
Consider a cooperative relay system that consists of $K + 1$ users, one acting as the source and $K$ serving as the relays. Each relay has two receive antennas. Let us denote the source by $s$, the destination by $d$, and label the relays from $1$ to $K$. $P_s$ is the source transmission power, ${h_{s,k,1}}$ and ${h_{s,k,2}}$ are the channel coefficients between source and first receiver antenna of relay  (i.e., the $s - k - 1$ link), and between source and the second receiver antenna of relay  (i.e., the $s - k - 2$ link), respectively. ${h_{k,d}}$ is the channel coefficient between relay $k$ and the destination (i.e., the $k - d$ link. $\sigma _k^2$ and $\sigma _d^2$ are noise variances at relay $k$ and destination, respectively. Moreover, the instantaneous SNR for $s - k - 1$, $s - k - 2$  and $k - d$ links are given by ${\gamma _{s,k,1}} = {P_s}{\left| {{h_{s,k,1}}} \right|^2}/\sigma _k^2$, ${\gamma _{s,k,2}} = {P_s}{\left| {{h_{s,k,2}}} \right|^2}/\sigma _k^2$  and ${\gamma _{k,d}} = {P_r}{\left| {{h_{k,d}}} \right|^2}/\sigma _d^2$.
In DF selection relaying, all relays attempt to decode the source’s message in phase I and act as candidate relays for selection in phase II only if it has successfully decoded the message. For simplicity, we assume the case where no diversity combining is employed at the destination. Hence, the system reduces to a dual-hop transmission where the maximum achievable rate is limited by the minimum capacity among  and  links. Given that relay  was selected, the end to end SNR can be computed as
\begin{align}
{\gamma _{SC}} = \mathop {\max }\limits_{k = 1,...,K} \min ({\gamma _{s,k}},{\gamma _{k,d}}){\rm{ }}
\end{align}
where ${\gamma _{s,k}} = \max \left( {{\gamma _{s,k,1}},{\gamma _{s,k,2}}} \right)$
\begin{figure}
  \centering
  \includegraphics[keepaspectratio,width=\columnwidth]{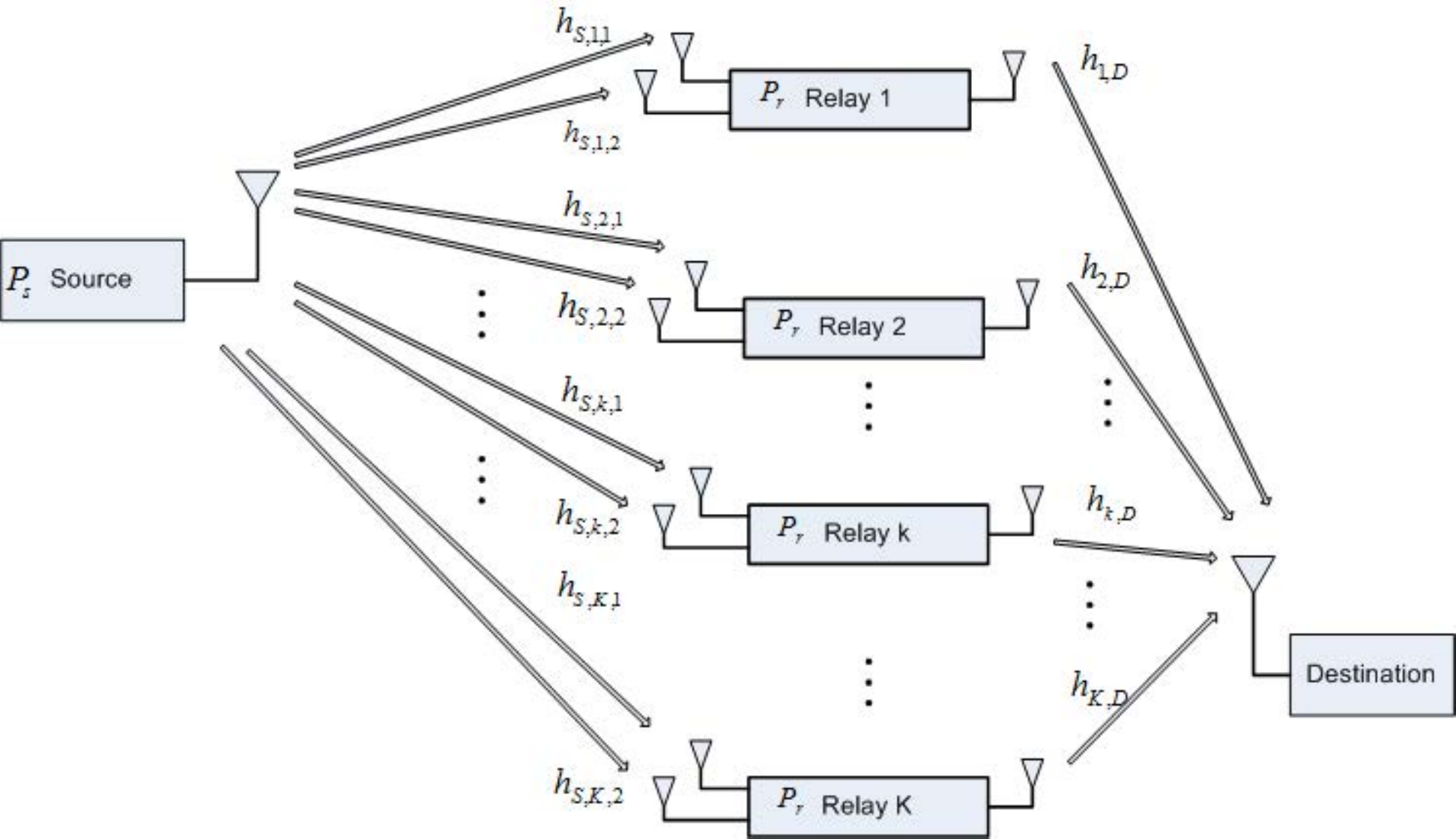}
  \caption{	System model of a Selective Combining cooperation.}
  \label{Fig1}
\end{figure}

\section{Performance Analysis}
The SNR per symbol of the this channel, $ \gamma  $ , is distributed according to a gamma distribution given by \citep{Simon}, where $m$ is Nakagami-$m$ fading parameters and $\overline \gamma  $  is average SNR per symbol. The Cumulative Distribution Function (CDF) of this channel are given by \citep{Ikki2007}, where $\Gamma (b,x)$ is the upper incomplete gamma function which is defined as $\Gamma (b,x) = \int\limits_x^\infty  {{e^{ - t}}{t^{b - 1}}} dt$  and   $\alpha  = \frac{m}{{\overline \gamma  }}$. Where ${\alpha _k} = \frac{{m{1_k}}}{{\Omega {1_k}}}$, ${\eta _k} = \frac{{m{2_k}}}{{\Omega {2_k}}}$ and ${\beta _k} = \frac{{m{3_k}}}{{\Omega {3_k}}}$ are the channel parameters for $s - k - 1$, $s - k - 2$ and $k - d$ links, respectively.
Since ${\gamma _{s,k}}$ and  ${\gamma _{k,d}}$ are independent gamma random variables and assuming that $m$ is integer and then by using  $\Gamma (n,x) = (n - 1)!{e^{ - x}}\sum\nolimits_{i = 0}^{n - 1} {\frac{{{x^i}}}{{i!}}}$ \citep{Speigel}, the CDF of $k$-th branch can be written as (see appendix)
\begin{align}
{P_{{\gamma _k}}}(\gamma ) =& 1 - \left[ {1 - {{\left\{ {1 - {e^{ - \alpha \gamma }}\sum\limits_{i = 0}^{m - 1} {\frac{{{\alpha ^i}{\gamma ^i}}}{{i!}}} } \right\}}^2}} \right]\left[ {{e^{ - \alpha \gamma }}\sum\limits_{l = 0}^{m - 1} {\frac{{{\alpha ^l}{\gamma ^l}}}{{l!}}} } \right]\nonumber\\
 =& 1 - 2{e^{ - 2\alpha \gamma }}\sum\limits_{i = 0}^{m - 1} {\sum\limits_{l = 0}^{m - 1} {\frac{{{\alpha ^{i + l}}{\gamma ^{i + l}}}}{{i!l!}}} }
{\rm{              }} + {e^{ - 3\alpha \gamma }}\sum\limits_{i = 0}^{m - 1} {\sum\limits_{j = 0}^{m - 1} {\sum\limits_{l = 0}^{m - 1} {\frac{{{\alpha ^{i + j + l}}{\gamma ^{i + j + l}}}}{{i!j!l!}}} } }
\end{align}

The Probability Density Function (PDF) of ${\gamma _k}$ can be computed by differentiating (1) with respect to ${\gamma}$ and after some simple manipulations as follows:
\begin{align}
{p_{{\gamma _k}}}(\gamma ) =& \frac{{4{e^{ - 2\alpha \gamma }}}}{{(m - 1)!}}\sum\limits_{u = 0}^{m - 1} {\frac{{{\alpha ^{m + u}}{\gamma ^{m - 1 + u}}}}{{u!}}} \nonumber\\
 -&\frac{{3{e^{ - 3\alpha \gamma }}}}{{(m - 1)!}}\sum\limits_{v = 0}^{m - 1} {\sum\limits_{w = 0}^{m - 1} {\frac{{{\alpha ^{m + v + w}}{\gamma ^{m - 1 + v + w}}}}{{v!w!}}} }
\end{align}
The PDF of selective combining SNR can be written as

\begin{align}
 {p_{{\gamma _{sc}}}} = K{\rm{ }}P_{{\gamma _k}}^{K - 1}\,{p_{{\gamma _k}}}(\gamma )                                                                            \end{align}

Using multinomial coefficient as (9) and after some manipulations, (8) can be simplified as (10)

 \begin{align}
 {(a + b + c)^{K - 1}} = \sum\limits_{p = 0}^{K - 1} {\sum\limits_{q = 0}^{K - 1 - p} {\frac{{K{\rm{!}}}}{{p!q!(K - 1 - p - q)!}}{{\rm{a}}^p}{{\rm{b}}^q}{{\rm{c}}^{K - 1 - p - q}}{\rm{   }}} }
  \end{align}

 \begin{align}
 {p_{{\gamma _{sc}}}} =& \sum\limits_{p = 0}^{{\rm{K - 1}}} \sum\limits_{q = 0}^{{\rm{K - 1}} - p} \Bigg\{
\frac{{{\rm{K!}}}}{{p!q!({\rm{K - 1}} - p - q)!}}
  \left( {{{\left( { - 2} \right)}^p}{e^{ - 2p\alpha \gamma }}{{\left[ {\sum\limits_{i = 0}^{m - 1} {\frac{{{\alpha ^i}{\gamma ^i}}}{{i!}}} } \right]}^{2p}}} \right)\nonumber\\\times&
  \left( {{e^{ - 3q\alpha \gamma }}{{\left[ {\sum\limits_{j = 0}^{m - 1} {\frac{{{\alpha ^j}{\gamma ^j}}}{{j!}}} } \right]}^{3q}}} \right)
 \Bigg\}  \times {p_{{\gamma _k}}}(\gamma )
\end{align}

Using multinomial coefficients and some simple manipulations, following relation is always held.

\begin{align}
{\left[ {\sum\limits_{i = 0}^{m - 1} {\frac{{{\alpha ^i}{\gamma ^i}}}{{i!}}} } \right]^{2p}} = \sum\limits_{{i_1} = 0}^{m - 1} {...\sum\limits_{{i_{2p}} = 0}^{m - 1} {\frac{{{{(\alpha \gamma )}^{\sum\limits_{t = 1}^{2p} {({i_t})} }}}}{{\prod\limits_{t = 1}^{2p} {{i_t}!} }}} }
\end{align}

Substituting (7) and (3) in (6) and after some simple manipulation the  can be derived by

\begin{align}
{p_{{\gamma _{SC}}}} = \widetilde \sum \sum\limits_{u = 0}^{m - 1} {\zeta {e^{ - \tau \gamma }}{\gamma ^\varphi }}  - \widetilde \sum \sum\limits_{v = 0}^{m - 1} {\sum\limits_{w = 0}^{m - 1} {\psi {\gamma ^\phi }{e^{ - \lambda \gamma }}} }
\end{align}

where

${\zeta _{p,q,u}} = \frac{4}{{u!(m - 1)!}}\frac{{{\alpha ^{\sum\limits_{t = 1}^{3q} {{j_t} + \sum\limits_{t = 1}^{2p} {{i_t}} }  + m + u}}}}{{\prod\limits_{t = 1}^{3q} {{j_t}!\prod\limits_{t = 1}^{2p} {{i_t}!} } }}\frac{{{\rm{K!}}{{( - 2)}^p}}}{{p!q!({\rm{K - 1}} - p - q)!}}$

${\tau _{p,q}} = (2(p + 1) + 3q)\alpha $

${\varphi _{p,q,u}} = \sum\limits_{t = 1}^{3q} {{j_t} + \sum\limits_{t = 1}^{2p} {{i_t} + m - 1 + u} } $

${\psi _{p,q,v,w}} = \frac{3}{{v!w!(m - 1)!}}\frac{{{\alpha ^{\sum\limits_{t = 1}^{3q} {{j_t} + \sum\limits_{t = 1}^{2p} {{i_t} + m + v + w} } }}}}{{\prod\limits_{t = 1}^{3q} {{j_t}!\prod\limits_{t = 1}^{2p} {{i_t}!} } }}\frac{{{\rm{K!}}{{( - 2)}^p}}}{{p!q!({\rm{K - 1}} - p - q)!}}$

${\lambda _{p,q}} = (2p + 3(q + 1))\alpha $

${\phi _{p,q,v}} = \sum\limits_{t = 1}^{3q} {{j_t} + \sum\limits_{t = 1}^{2p} {{i_t}} }  + m - 1 + v + w$

$\widetilde \sum  = \sum\limits_{p = 0}^{{\rm{K - 1}}} {\sum\limits_{q = 0}^{{\rm{K - 1}} - p} {\sum\limits_{{i_1} = 0}^{m - 1} {...\sum\limits_{{i_{2p}} = 0}^{m - 1} {\sum\limits_{{j_1} = 0}^{m - 1} {...\sum\limits_{{j_{3q}} = 0}^{m - 1} {} } } } } } $

After taking the Laplace transform of the PDF and some simple algebraic manipulations, the MGF can be determined by

 \begin{align}
{M_{{\gamma _{sc}}}}(s) = \widetilde \sum \sum\limits_{u = 0}^{m - 1} {\frac{{{\zeta _u}{\varphi _u}!}}{{{{(\tau  + s)}^{{\varphi _u} + 1}}}}}  - \widetilde \sum \sum\limits_{v = 0}^{m - 1} {\sum\limits_{w = 0}^{m - 1} {\frac{{{\psi _{v,w}}{\phi _{v,w}}!}}{{{{(\lambda  + s)}^{{\phi _{v,w}} + 1}}}}} }
\end{align}

The symbol error probability (SEP) can be derived in closed-form using the MGF in (9). For M-PSK modulation, the SER is given by \citep{Simon} and approximation is given by \citep{PIMRC}.
The channel capacity can be computed as \citep{Ikki}

 \begin{align}
\overline C  = \frac{{BW}}{2}\int\limits_0^\infty  {{{\log }_2}(1 + \gamma ){p_\gamma }(\gamma )d\gamma }
\end{align}

where $BW$ is the transmitted signal bandwidth. By substituting (8) in (10) and using \citep{Prudnikov}, the averaged channel capacity can be obtained in closed-form in (11)

\begin{align}
\overline C  = \frac{{BW}}{{2\ln 2}}\left\{ \begin{array}{l}
\widetilde \sum \sum\limits_{u = 0}^{m - 1} {{\zeta _u}{\tau ^{ - {\varphi _u} - 1}}G_{3,2}^{1,3}\left( {\left. {\frac{1}{\tau }} \right|_{1,0}^{ - {\varphi _u},1,1}} \right)}  - \\
\widetilde \sum \sum\limits_{v = 0}^{m - 1} {\sum\limits_{w = 0}^{m - 1} {{\psi _{v,w}}{\eta ^{ - {\phi _{v,w}} - 1}}G_{3,2}^{1,3}\left( {\left. {\frac{1}{\eta }} \right|_{1,0}^{ - {\phi _{v,w}},1,1}} \right)} }
\end{array} \right\}
\end{align}

where $G_{p,q}^{m,n}\left( {\left. . \right|_{({b_p})}^{({a_p})}} \right)$ is the Meijer-G function \citep{Prudnikov}.

\section{Optimization Problem}
To minimize the outage probability, we define optimization problem as follows:

\begin{align}
&\mathop {\min }\limits_{{P_s},{P_r}} {P_{out}}\nonumber\\
&s.t.{\rm{ }}{{\rm{P}}_s} + {P_r} = {P_{tot}}\nonumber\\
&{P_s} > 0,{\rm{ }}{{\rm{P}}_r} > 0
\end{align}

Where ${P_{tot}}$ is the total power of the system. Because of non-convexity of the above problem \citep{Boyd}, we solved this problem numerically for Nakagami-$m$ fading channels. We have compared the optimal power allocation with adaptive and equal power allocation. In adaptive power allocation, source and relay powers are allocated as follows:

\begin{align}
{P_s} =& \left( {\frac{1}{{Ant + 1}} + \frac{{\Omega {3_k}}}{{{\Omega _k} + \Omega {3_k}}} + \frac{{{m_k}}}{{{m_k} + m{3_k}}}} \right)\frac{{{P_{tot}}}}{3}\\
{P_r} =& \left( {\frac{{Ant}}{{Ant + 1}} + \frac{{{\Omega _k}}}{{{\Omega _k} + \Omega {3_k}}} + \frac{{m{3_k}}}{{{m_k} + m{3_k}}}} \right)\frac{{{P_{tot}}}}{3}
  \end{align}

Where $Ant$ is the number of antennas, ${m_k} = m{1_k} = m{2_k}$ and $\Omega {1_k} = \Omega {2_k} = {\Omega _k}$.

For the Rayleigh case, The problem can be formulated as follows:

\begin{align}
&\mathop {\min }\limits_{{P_s},{P_r}} \prod\limits_{k = 1}^K {{P_{{\gamma _k}}}}  \equiv \mathop {\min }\limits_{{P_s},{P_r}} \sum\limits_{k = 1}^K {\log ({P_{{\gamma _k}}})} \nonumber\\
&s.t.{\rm{ }}{{\rm{P}}_s} + {P_r} = {P_{tot}},{P_s} > 0,{\rm{ }}{{\rm{P}}_r} > 0
\end{align}

By approximating the CDF of $k$-th branch, then we have \citep{Wolfram}

\begin{align}
{P_{out}} = \prod\limits_{k = 1}^K {\left( {{\beta _k}\gamma  + {\alpha _k}{\eta _k}{\gamma ^2} - {\alpha _k}{\eta _k}{\beta _k}{\gamma ^3}} \right)}
\end{align}

We can rewrite problem as follow:

\begin{align}
&\mathop {\min }\limits_{{P_s},{P_r}} \sum\limits_{k = 1}^K {\log } \left( {\frac{{{a_k}}}{{{P_r}}} + \frac{{{b_k}}}{{P_s^2}} - \frac{{{c_k}}}{{P_s^2{P_r}}}} \right)\nonumber\\
&s.t.{\rm{ }}{{\rm{P}}_s} + {P_r} = {P_{tot}}\nonumber\\
&{P_s} > 0,{\rm{ }}{{\rm{P}}_r} > 0
\end{align}

where ${a_k} = \sigma _d^2{\gamma _{th}}/\mu _{k,d}^2$, ${b_k} = \sigma _k^4\gamma _{th}^2/\mu _{s,k,1}^2\mu _{s,k,2}^2$ and ${c_k} = \sigma _k^4\sigma _d^2\gamma _{th}^3/\mu _{s,k,1}^2\mu _{s,k,2}^2\mu _{k,d}^2$.
${\gamma _{s,k,1}}$, ${\gamma _{s,k,2}}$ and  ${\gamma _{k,d}}$ are exponentially distributed with means $\Omega {1_k} = {P_s}\mu _{s,k,1}^2/\sigma _k^2$, $\Omega {2_k} = {P_s}\mu _{s,k,2}^2/\sigma _k^2$ and $\Omega {3_k} = {P_s}\mu _{k,d}^2/\sigma _d^2$, respectively.
By introducing the Lagrange multiplier and by applying the KKT conditions \citep{Boyd}, after some simple manipulations, the power allocation equation can be obtained as

\begin{align}
P_s^3 + {h_1}P_s^2 + {h_2}{P_s} + {h_3} = 0
\end{align}

where

\begin{align}
{h_1} =  - \frac{{2{b_k}}}{{{a_k}}},{h_2} = \frac{{4{b_k}{P_{tot}} - 3{c_k}}}{{{a_k}}},{h_3} = \frac{{2{c_k}{P_{tot}} - 2{b_k}P_{tot}^2}}{{{a_k}}}
\end{align}
Using \citep{Speigel} and after some simple manipulation, the optimal power allocation can be obtained as

\begin{align}
&{P_s} = \left\{ \begin{array}{l}
S + T - \frac{{{h_1}}}{3}{\rm{                         D}} \ge {\rm{0}}\\
2\sqrt { - Q} \cos (\frac{{\theta  + 4\pi }}{3}) - \frac{{{h_1}}}{3}{\rm{    D  <  0}}
\end{array} \right.\nonumber\\
&{{\mathop{\rm P}\nolimits} _r} = {P_{tot}} - {P_s}
\end{align}

where

$Q = \frac{{3{h_2} - h_1^2}}{9},R = \frac{{9{h_1}{h_2} - 27{h_3} - 2h_1^3}}{{54}}$,

$S = \sqrt[3]{{R + \sqrt D }}$,

$T = \sqrt[3]{{R - \sqrt D }}$,

$D = {Q^3} + {R^2}$,

$\cos (\theta ) = \frac{R}{{\sqrt { - {Q^3}} }}$

\section{Simulation Results}
 To verify our analytical results, we show statistical simulation results and compare them with our analytical for the outage probability, symbol error probability and average channel capacity.
In all simulation and analytical results we consider the system with 16PSK modulation and  ${\gamma _{th}} = 3$ for two topologies:

1) Symmetric case e.g.,

$\Omega {1_k} = \Omega {2_k} = \Omega {3_k} = 3,{\rm{ m}}{{\rm{1}}_k} = {\rm{m}}{{\rm{2}}_k} = {\rm{m}}{{\rm{3}}_k} = 2$

and 2) Asymmetric case e.g.,

$\Omega {1_1} = \Omega {2_1} = {\rm{m}}{{\rm{1}}_1} = {\rm{m}}{{\rm{2}}_1} = 1,{\rm{ }}\Omega {1_2} = \Omega {2_2} = {\rm{m}}{{\rm{1}}_2} = {\rm{m}}{{\rm{2}}_2} = 2,{\rm{ }}\Omega {1_3} = \Omega {2_3} = {\rm{m}}{{\rm{1}}_3} = {\rm{m}}{{\rm{2}}_3} = 3,{\rm{ }}$

 $\Omega {3_1} = {\rm{m}}{{\rm{3}}_1} = 3,{\rm{ }}\Omega {3_2} = {\rm{m}}{{\rm{3}}_2} = 2,{\rm{ }}\Omega {3_3} = {\rm{m}}{{\rm{3}}_3} = 1$.

In Fig. 2, and Fig. 3, the outage and symbol error probability for DF relay selection scheme versus SNR for Nakagami-$m$ fading channels are shown. As we can see, our proposed approach has more than 1 dB performance in comparison with \citep{Duong:2009:Comletter}.

\begin{figure}
  \centering
  \includegraphics[keepaspectratio,width=\columnwidth]{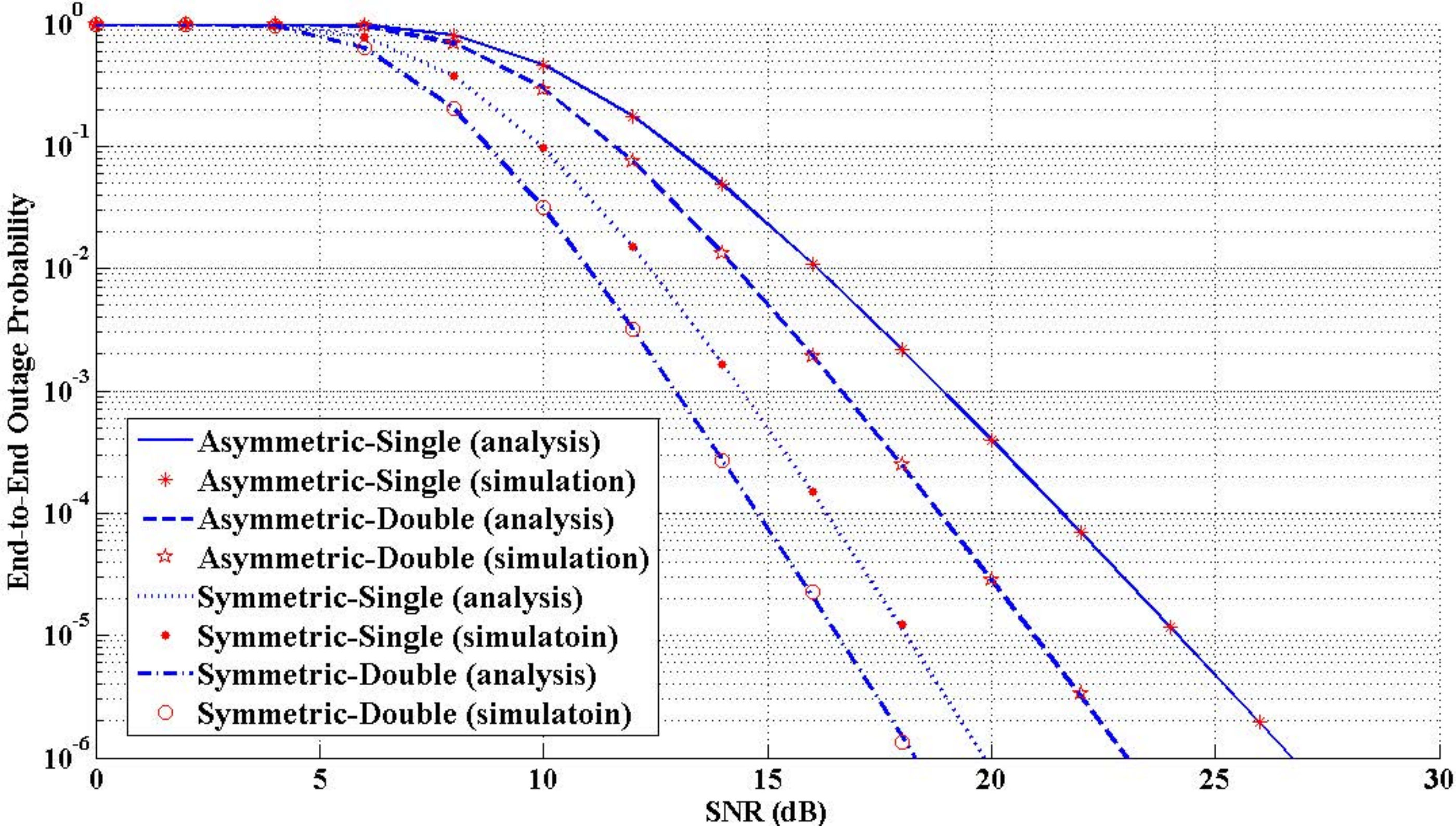}
  \caption{	Outage probability of the selective combining DF relay system.}
  \label{Fig2}
\end{figure}

\begin{figure}
  \centering
  \includegraphics[keepaspectratio,width=\columnwidth]{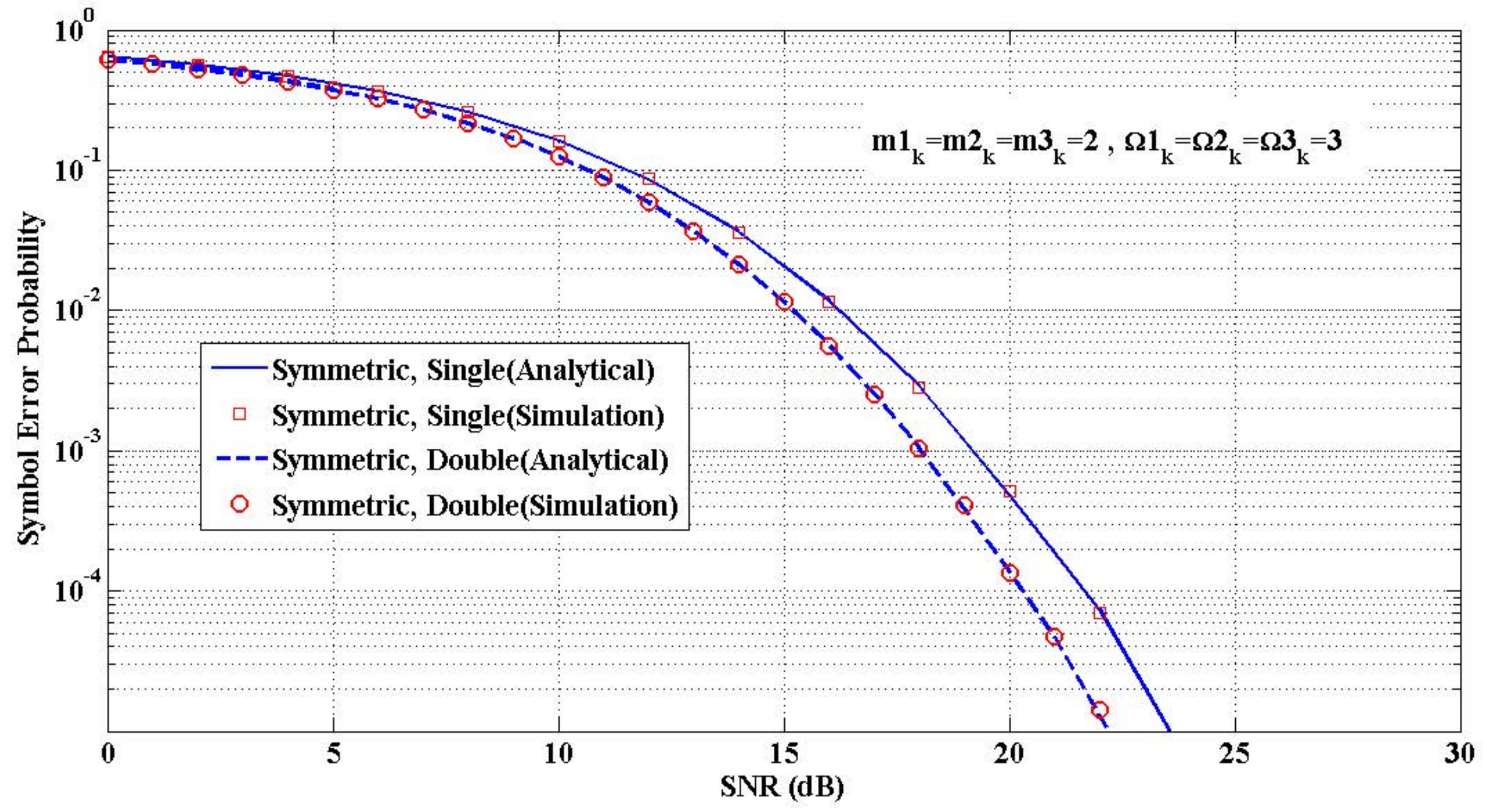}
  \caption{	Symbol error probability of the selective combining DF relay system}
  \label{Fig3}
\end{figure}

In Fig. 4, the outage probability versus SNR in different number of relays and different number of relay antennas are shown. As we can see, as the number of relays increase, the double antennas case has more advantage versus single antenna case. Also, using 4 relays with two antennas has more advantage than using 5 relays with single antenna in considerable range of SNR.
It was shown that the average SNR gain increases with the number of the diversity combining antennas, but not linearly. The largest gain increment is achieved by going from one receiver antenna (no diversity) to two antennas. Increasing the number of receiving antenna from two to three will give much less gain than going from one to two \citep{Halim:2007:WCOM}. Therefore, in this paper we studied a relay system with two receiving antennas in order to have the largest gain increment and the least complexity increment.

\begin{figure}
  \centering
  \includegraphics[keepaspectratio,width=\columnwidth]{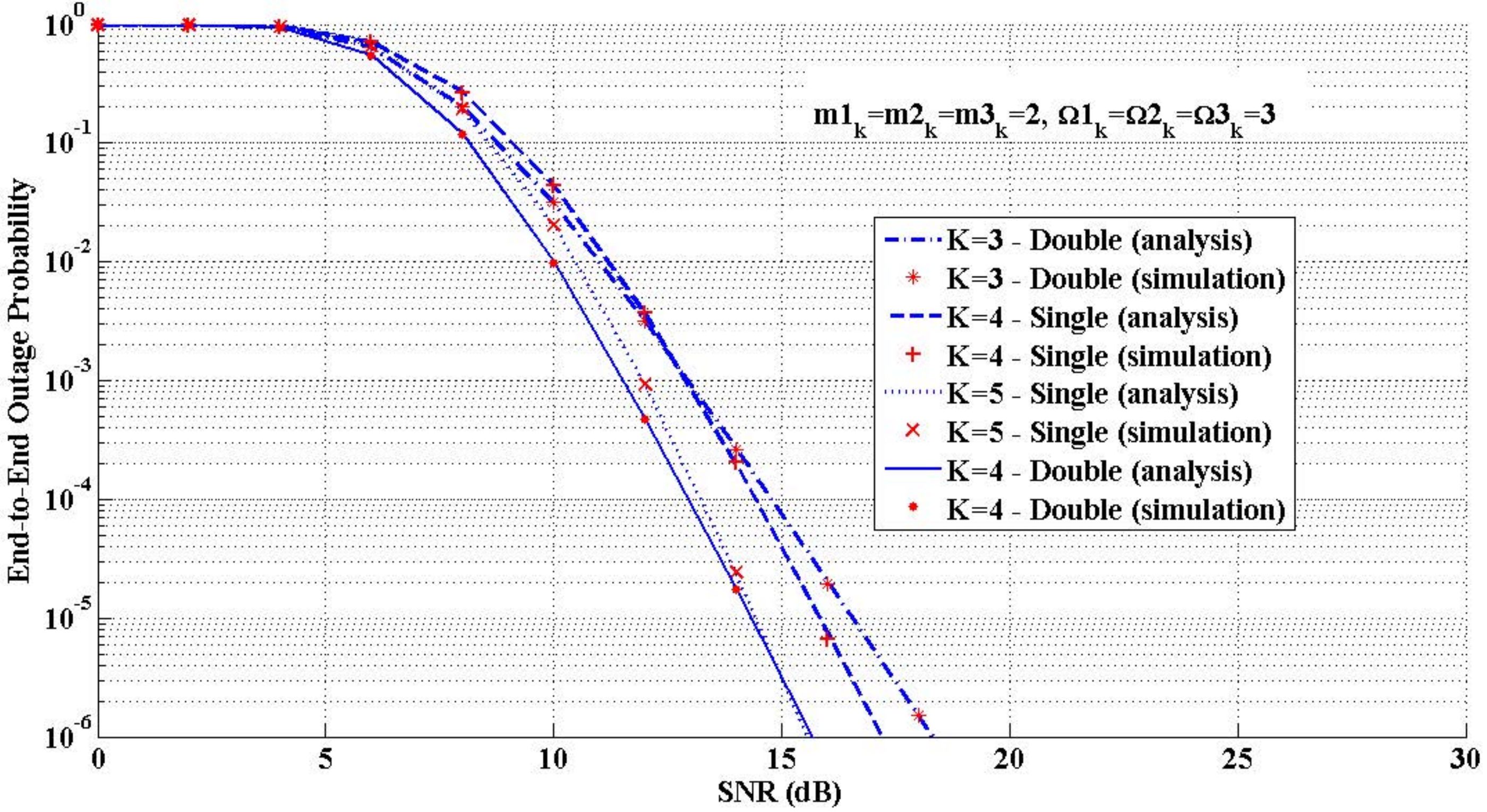}
  \caption{	Outage probability versus SNR in different number of relays relay antennas.}
  \label{Fig4}
\end{figure}

 In Fig. 5 and 6, we have shown the outage probability of the SC-DF relay for Nakagami-$m$ and Rayleigh fading channels, respectively. As we can see, the optimal power allocation has more than 1dB performance in comparison with equal power allocation. Also, approximate optimal power allocation is very close to optimal numerical optimization.

\begin{figure}
  \centering
  \includegraphics[keepaspectratio,width=\columnwidth]{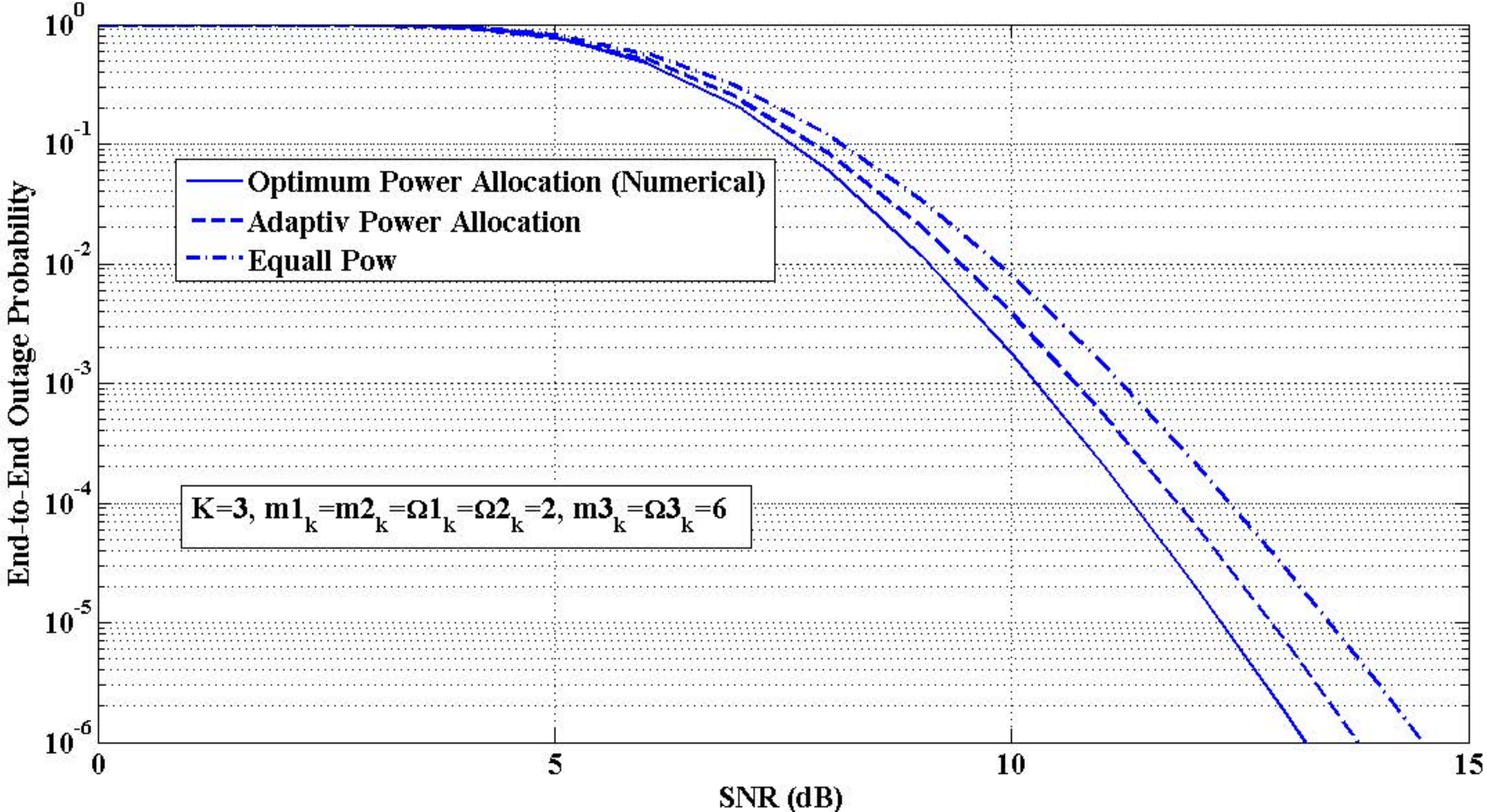}
  \caption{	Outage probability of the SC-DF relay system (Optimal and Equal power allocation).}
  \label{Fig5}
\end{figure}

\begin{figure}
  \centering
  \includegraphics[keepaspectratio,width=\columnwidth]{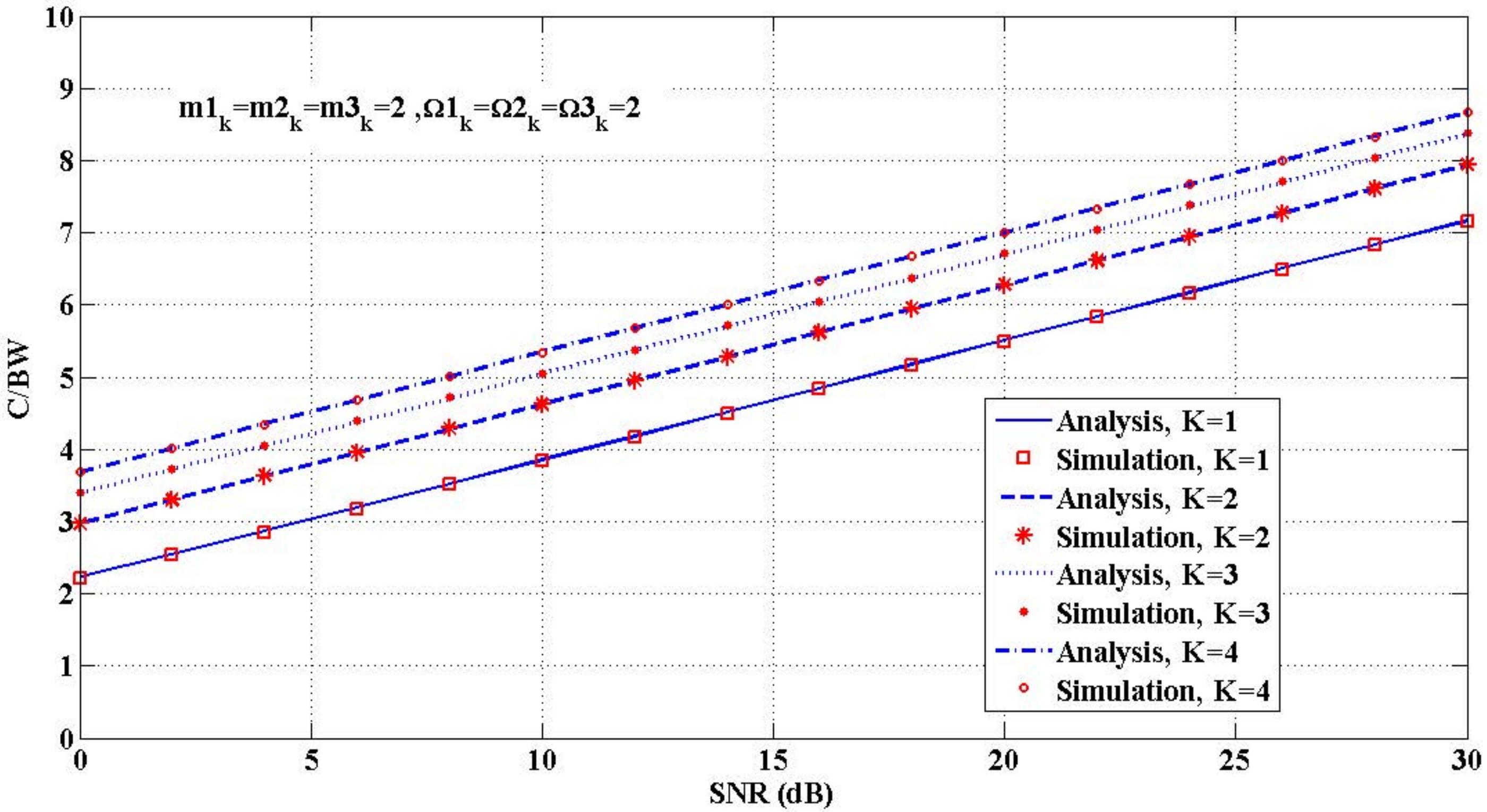}
  \caption{	Outage probability of the SC-DF relay system (Rayleigh fading).}
  \label{Fig6}
\end{figure}

   In Fig. 7, the average channel capacity in analytical and simulation for Nakagami-$m$ fading channels has been shown. Where $K$ is the number of relays. As we can see, the average channel capacity increases as $K$ increases. Moreover, SC scheme can outperform the direct transmission at higher SNR if the total number of relays ($K$)  can be increased.

\begin{figure}
  \centering
  \includegraphics[keepaspectratio,width=\columnwidth]{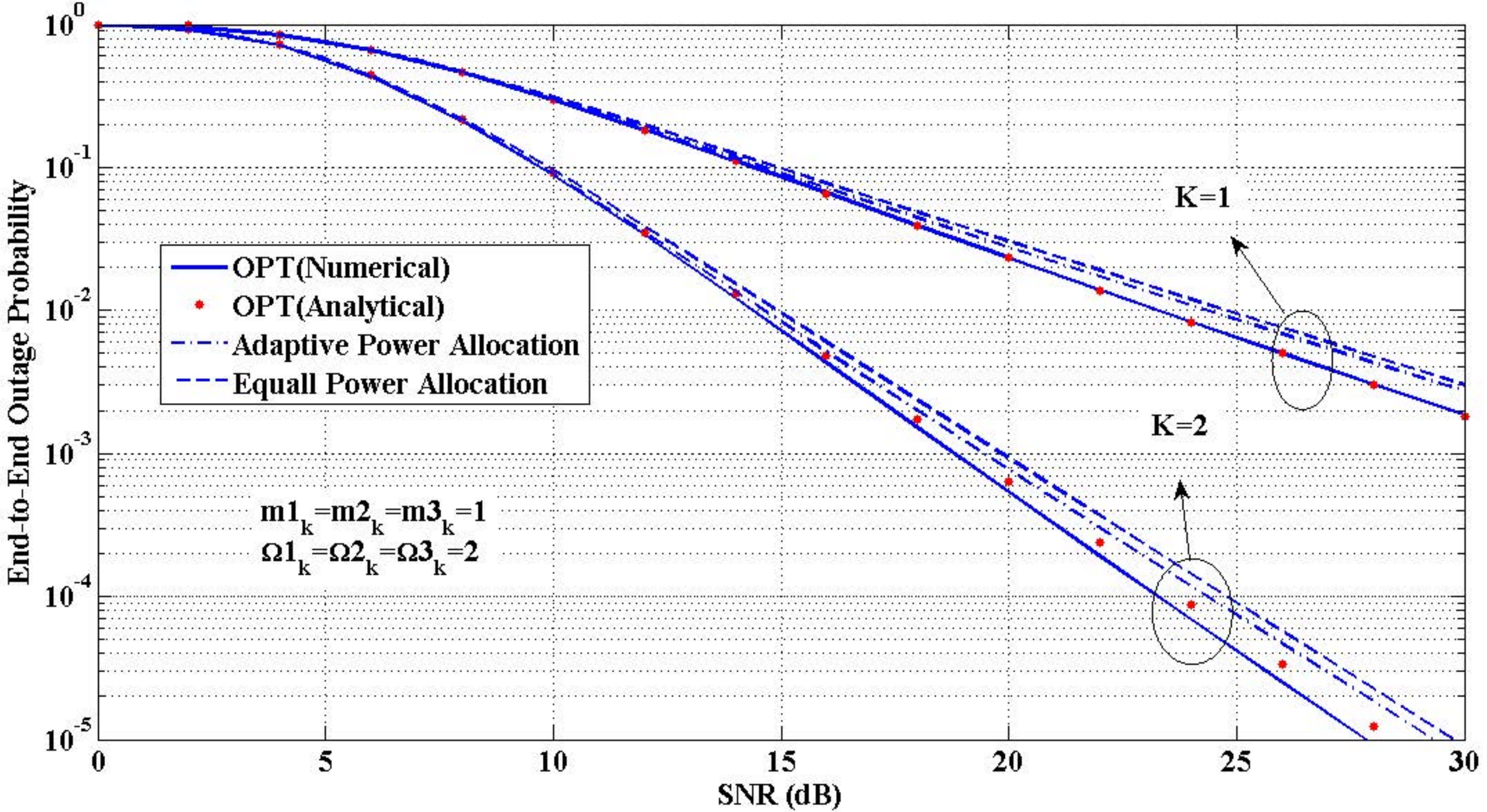}
  \caption{	Average channel capacity for the SC-DF relay system (with and without direct link).}
  \label{Fig7}
\end{figure}

\section{Future Research Directions}
There are other channels that can be analyzed under this strategy such as Log-normal, Rican (Nakagami-$n$) and Hoyt (Nakagami-$q$) fading channels. From the practical point of view, Log-normal distribution is encountered in many communication scenarios. For instance, when indoor communication is used at which users are moving, Log-normal distribution not only models the moving objects, but also the reflection of the bodies. Moreover, it models the action of communicating with robots in a closed environment like a factory.

In indoor radio propagation environments, terminals with low mobility have to rely on macroscopic diversity to overcome the shadowing from indoor obstacles and moving human bodies. Indeed, in such slowly varying channels, the small-scale and large-scale effects tend to get mixed. In this case, Log-normal statistics accurately describe the distribution of the channel path gain. In most of the scenarios, we assume that the relay and the destination have perfect knowledge of the channel gains; however, this is not practical. This scenario can be reconsidered by assuming that we do not have the perfect knowledge of the channel gains between the source-relay and the relay-destination. Another important relaying protocol that has not been considered in this chapter is the AF relaying. As we said, in this method there is an analog amplifier in the relay witch amplifies the received signal and then forwards it, and no digital processing such as coding and error estimation takes effect, so implementing AF relay based network is easy. Most importantly, if our area of work is in the high SNR, both AF and DF have the same performance, so we’d better use the AF relaying. One of other important drawbacks of the wireless communication systems is the delay feedback in which the selected relay index is sent from the destination to the relays.

Till we have slow varying channel compared to the time slot and the delay feedback, there will be no serious problem for the performance of the system; however, if the channel is varying fast, the channel gain of the selected relay will change, and the information will not be sent through the best channel.
Interested readers can refer to \citep{Aalo,5452964,5491390,5706376,4917860,4801499,5089520,Alouini1,4205088,1580784,1611050,5199262,4786407,soleimaniICEE,soleimaniTCOM,soleimaniTVT,soleimaniWPC,soleimaniWCMC,soleimaniAEU,soleimaniglobecom,soleimaniicassp,soleimaniicc,soleimaniist,soleimaniISCC} and references therein.
 \section{Conclusion}\label{sec:Conclusion}
In this chapter, we have analyzed the performance of a multi-antenna selective decode-and-forward
relaying system over Nakagami-$m$ fading channels without and with the direct link between source and
destination. We have derived the outage probability, moment generation function, symbol error
probability and average channel capacity in closed-form. Also, we derived an approximation problem to
minimize outage probability and then we solved the problem. Simulation results verified the accuracy and
the correctness of the proposed analysis. The complexity of double antenna case versus single antenna
case, isn’t very high and instead of increasing the number of relays, increasing the number of antennas is
a practical option.

\end{document}